\documentclass[fleqn,usenatbib]{mnras}

\usepackage{newtxtext,newtxmath}
\usepackage[T1]{fontenc}

\DeclareRobustCommand{\VAN}[3]{#2}
\let\VANthebibliography\thebibliography
\def\thebibliography{\DeclareRobustCommand{\VAN}[3]{##3}\VANthebibliography}

\usepackage{graphicx}	
\usepackage{amsmath}	
\usepackage{lipsum}
\usepackage{subcaption}
\usepackage{xspace}
\usepackage{xfrac}
\usepackage{soul}

\newcommand{\bU}{\langle bU\rangle}
\newcommand{\bM}{\langle b_m\rangle}
\newcommand{\bPe}{\langle bP_e\rangle}
\newcommand{\brsfr}{\langle b\rho_{\rm SFR}\rangle}
\newcommand{\rsfr}{\rho_{\rm SFR}}
\newcommand{\nv}{\hat{\bf n}}
\newcommand{\hMpc}{h^{-1}\,{\rm Mpc}}
\def\flamingo{\texttt{FLAMINGO}\xspace}
\newcommand{\bemu}{{\tt Baccoemu}\xspace}
\newcommand{\hfit}{{\tt HALOFIT}\xspace}
\newcommand{\ngal}[1]{\bar{n}=10^{- #1}\,h^3\,{\rm Mpc}^{-3}}
\newcommand{\ngall}{\bar{n} = \{10^{-2}, 10^{-3}, 10^{-4}\}\,h^3 \text{Mpc}^{-3}}

\title[The impact of galaxy bias on cross-correlation tomography]{The impact of galaxy bias on cross-correlation tomography}

\author[S. Maleubre et al.]{
Sara Maleubre,$^{1}$\thanks{E-mail: sara.maleubremolinero@physics.ox.ac.uk}
Matteo Zennaro,$^{1}$
David Alonso,$^{1}$
Ian G McCarthy,$^{2}$
Matthieu Schaller,$^{3,4}$
and Joop Schaye$^{3}$
\\
$^1$Department of Physics, University of Oxford, Denys Wilkinson Building, Keble Road, Oxford OX1 3RH, United Kingdom\\
$^{2}$Astrophysics Research Institute, Liverpool John Moores University, 146 Brownlow Hill, Liverpool L3 5RF, United Kingdom\\
$^{3}$Leiden Observatory, Leiden University, PO Box 9513, NL-2300 RA Leiden, the Netherlands\\
$^{4}$Lorentz Institute for Theoretical Physics, Leiden University, PO box 9506, 2300 RA Leiden, the Netherlands 
}

\date{Accepted XXX. Received YYY; in original form ZZZ}

\pubyear{2025}

\begin{document}
\label{firstpage}
\pagerange{\pageref{firstpage}--\pageref{lastpage}}
\maketitle

\begin{abstract}
  The cross-correlation of galaxies at different redshifts with other tracers of the large-scale structure can be used to reconstruct the cosmic mean of key physical quantities, and their evolution over billions of years, at high precision. However, a correct interpretation of these measurements must ensure that they are independent of the clustering properties of the galaxy sample used. In this paper we explore different prescriptions to extract tomographic reconstruction measurements and use the FLAMINGO hydrodynamic simulations to show that a robust estimator, independent of the small-scale galaxy bias, can be constructed. We focus on the tomographic reconstruction of the halo bias-weighted electron pressure $\langle bP_e\rangle$ and star-formation density $\langle b\rho_{\rm SFR}\rangle$, which can be reconstructed from tomographic analysis of Sunyaev-Zel'dovich and cosmic infrared background maps, respectively. We show that these quantities can be reconstructed with an accuracy of 1-3\% over a wide range of redshifts, using different galaxy samples. We also show that these measurements can be accurately interpreted using the halo model, assuming a sufficiently reliable model can be constructed for the halo mass function, large-scale halo bias, and for the dependence of the physical quantities being reconstructed on halo mass.
\end{abstract}





\section{Introduction}\label{sec:intro}
  Galaxies are biased tracers of the underlying fluctuations in the matter density. In general, the abundance of galaxies of a given type at different points in space is strongly correlated with the density of matter and, on sufficiently large scales, both overdensities can be linearly related via
  \begin{equation}
    \delta_g\simeq b_g\,\delta_m,
  \end{equation}
  where $b_g$ is the linear galaxy bias parameter, and $\delta_g$ and $\delta_m$ are the galaxy and matter overdensities. On small scales, however, the galaxy bias relation becomes significantly more complicated and difficult to model. Non-linear, non-local, and stochastic contributions become important, and their impact on galaxy clustering statistics is difficult to predict with high accuracy \citep{1611.09787,2110.05408,2112.00012}.

  Galaxies are also \emph{local} large-scale structure (LSS) tracers: their abundance at a given redshift tracks the matter overdensity, as well as any other physical quantity correlated with it, at the same redshift. This makes it possible to use the cross-correlation of galaxies with different projected tracers of the matter fluctuations to reconstruct the redshift dependence of interesting quantities, in a technique commonly called ``tomography'' or ``tomographic reconstruction'' \citep{0805.1409,1302.0857,1810.00885}. In broad terms, the large-scale amplitude of the cross-correlation between galaxies and a given LSS tracer $U$ depends on the bias-weighted average of $U$, $\bU$\footnote{Note that this ``bias weighting'' refers to the bias of the dark matter haloes populated by $U$, and not to the bias of the galaxies used in the cross-correlation.}, and on the clustering properties of the galaxies used in the analysis. The auto-correlation of these galaxies can then be used to constrain the galaxy bias relation, resulting in a measurement of $\bU$. 
  
  This technique has been widely used for multiple science purposes. In the ``CMB lensing tomography'' approach, cross-correlations with maps of the Cosmic Microwave Background lensing convergence have been used to reconstruct the growth of structure as a function of time \citep{2105.03421,2111.09898,2402.05761,2410.10808}. Cross-correlations with maps of the thermal Sunyaev-Zel'dovich effect (tSZ) have been used to reconstruct the bias-weighted mean electron pressure $\bPe$, which can be used to study the thermodynamics of the intergalactic medium, as well as the amplitude of matter fluctuations \citep{1608.04160,1904.13347,1909.09102,2006.14650,2201.12591,2210.08633}. The same approach can also be used to measure the bias-weighted star formation rate density $\brsfr$, as shown by \citep{2204.01649,2206.15394,2310.10848,2504.05384}. Cross-correlation with intensity maps at different wavelengths have also been used to reconstruct other key physical quantities, including the radio, ultraviolet, and gamma-ray background \citep{1810.00885,2307.14881,2311.17641}. Finally, the same principle is behind the so-called ``clustering redshifts'' approach: cross-correlations between galaxy samples with known redshifts, and samples with uncertain redshifts can be used to reconstruct the redshift distribution of the latter \citep{0805.1409,1003.0687,1302.0857,1303.0292,1303.4722}.

  In spite of the broad applicability of tomographic reconstruction, and its ability to recover model-independent measurements of interesting physical quantities, two outstanding questions have limited its use in cosmological analysis:
  \begin{itemize}
    \item Are the estimated values of $\bU$ unbiased, and independent of the clustering properties of the galaxy sample used to measure it?
    \item Can the measured values of $\bU$ be interpreted from more fundamental descriptions (e.g. through the halo model)?
  \end{itemize}
  Answering these questions is important for several reasons. First, past efforts at tomographic reconstruction have often made use of relatively small scales in order to boost the sensitivity of the resulting measurements \citep{2504.05384} (this is particularly common in the case of clustering redshifts \citep{2006.14650,2012.08569}). However, on these scales, galaxy bias can be significantly scale-dependent, affecting the shape of galaxy auto- and cross-correlations in different ways, and potentially biasing the recovered quantities. Second, an unbiased measurement of e.g. $\bPe$ is not immediately useful unless theoretical predictions can be built for it in terms of both cosmology and gas thermodynamics. Although this has been addressed in e.g. \cite{Young2021} in the particular case of tSZ tomography, a more general study has not been carried out yet.
  
  This paper addresses these two questions, finding that they can both be answered in the affirmative assuming suitable estimators and scale ranges are used. To do so, we will select different galaxy samples within the \flamingo suite of hydrodynamical simulations, taking advantage of their large simulation volume to probe the realistic correlation between matter, galaxies, and different baryonic properties over a wide range of linear and non-linear scales. We will apply a tomographic reconstruction estimator to all of these samples, using their auto-correlation and cross-correlation with various LSS fields $U$. In particular, we will target the matter overdensity, the thermal gas pressure, and the star formation-rate density. We will then assess the ability of the estimator to recover consistent and unbiased measurements of $\bU$, and the possibility of interpreting the measurements in the context of the halo model.

  This paper is structured as follows. Section \ref{sec:theory} introduces the theoretical background of tomographic reconstruction. The simulations used, including the different galaxy samples and LSS quantities extracted from them, are described in Section \ref{sec:sims}. Our results are presented in Section \ref{sec:res}, including an assessment of the robustness of $\bU$ measurements to small-scale galaxy bias, and their theoretical interpretation. Finally, Section \ref{sec:conc} summarises our findings.


\section{Tomography and galaxy bias}\label{sec:theory}
  \subsection{Cross-correlation-based tomography}\label{ssec:theory.tomo}
    Consider a projected field $u(\nv)$ defined on the celestial sphere, where $\nv$ is a directional unit vector, tracing a three-dimensional physical quantity $U({\bf x},z)$\footnote{Note that we use redshift $z$ as an effective time variable.} through a projection integral of the form
    \begin{equation}
      u(\nv)=\int d\chi\,q_u(\chi)\,U(\chi\nv,z(\chi)),
    \end{equation}
    where $q_u(\chi)$ is a known radial (comoving distance) kernel. Multiple examples of such fields exist in cosmology and astrophysics. Here we will consider the following:
    \begin{itemize}
      \item The {\bf CMB lensing} convergence field \citep{astro-ph/0601594} is a projected tracer of the matter overdensity ($U=\delta_m$):
      \begin{equation}
        \kappa(\nv)=\int d\chi\,q_\kappa(\chi)\,\delta_m(\chi\nv,z(\chi)),
      \end{equation}
      where the lensing kernel is
      \begin{equation}
        q_\chi(\chi)\equiv\frac{3}{2}H_0^2\Omega_m\,(1+z)\chi\frac{\chi_{\rm LSS}-\chi}{\chi_{\rm LSS}}.
      \end{equation}
      Here $\Omega_m$ and $H_0$ are the cosmic non-relativistic matter fraction and the Hubble parameter, respectively, and $\chi_{\rm LSS}$ is the comoving distance to the last-scattering surface.
      \item The {\bf thermal Sunyaev-Zel'dovich} effect \citep{astro-ph/0208192} Compton-$y$ parameter is a projected tracer of the thermal electron pressure $P_e$:
      \begin{equation}
        y(\nv)=\int d\chi\,q_y(\chi)\,P_e(\chi\nv,z(\chi)),\hspace{12pt}q_y(\chi)\equiv\frac{\sigma_T}{m_ec^2}\frac{1}{1+z},
      \end{equation}
      where $\sigma_T$ and $m_e$ are the Thomson scattering cross section and the electron mass, respectively, and $c$ is the speed of light.
      \item The {\bf Cosmic Infrared Background} intensity at a given observed frequency $I_\nu$ is sourced by dust emission in star-forming galaxies, and is thus a tracer of the star formation-rate (SFR) density $\rsfr$ \citep{2006.16329}:
      \begin{equation}
        I_\nu(\nv)=\int d\chi\,q_{{\rm CIB},\nu}(\chi)\,\rsfr(\chi\nv,z(\chi)),
      \end{equation}
      where the radial kernel is
      \begin{equation}
        q_{{\rm CIB},\nu}(\chi)\equiv\frac{\chi^2\,S^{\rm eff}_\nu(z)}{K}.
      \end{equation}
      Here $K$ is the calibration constant relating far infrared luminosity and SFR (sometimes called the ``Kennicut constant'' \citep{astro-ph/9807187}), and $S_\nu^{\rm eff}(z)$ is the mean flux of infrared sources at redshift $z$ normalised to unit luminosity \citep[see e.g.][]{2006.16329}.
    \end{itemize}

    In addition to these, consider the projected overdensity of galaxies
    \begin{equation}
      \Delta_g(\nv)=\int d\chi\,q_g(\chi)\,\delta_g(\chi\nv,z(\chi)),
      \hspace{12pt}q_g(\chi)\equiv\frac{H(z)}{c}\,p(z)
    \end{equation}
    where $\delta_g$ is the 3D galaxy overdensity, $H(z)$ is the expansion rate at redshift $z$, and $p(z)$ is the redshift distribution of the galaxies. It is often possible to obtain reasonably accurate measurements for galaxy redshifts, and thus well-localised samples can be selected, in which case $p(z)$ is a compact function centered at a mean redshift $\bar{z}_g$.

    The angular power spectrum between any two projected quantities, $u$ and $v$, is given by
    \begin{equation}
      C_\ell^{uv}=\int \frac{d\chi}{\chi^2}\,q_u(\chi)\,q_v(\chi)\,P_{UV}\left(k=\frac{\ell+1/2}{\chi},z(\chi)\right),
    \end{equation}
    where $P_{UV}(k,z)$ is the three-dimensional power spectrum between the two 3D quantities, $U$ and $V$, that the projected quantities trace.
    
    The halo model \citep{astro-ph/0001493,astro-ph/0005010,astro-ph/0206508} provides a useful framework to make predictions for the power spectrum of general LSS tracers. A key prediction of this model is that, on sufficiently large scales, the power spectrum $P_{UV}(k,z)$ can be approximated as
    \begin{equation}\label{eq:pk_hm}
      P_{UV}(k,z)\simeq \langle bU(z)\rangle\,\langle bV(z)\rangle\,P_{mm}(k,z),
    \end{equation}
    where $P_{mm}(k,z)$ is the matter power spectrum, and we have defined the bias-weighted average:
    \begin{equation}\label{eq:bU}
      \bU = \int dM\,n(M)\,b_h(M)\,\tilde{U}(M),
    \end{equation}
    where $n(M)$ is the halo mass function, $b_h(M)$ is the linear halo bias, and $\tilde{U}(M)$ is the average integral of the halo profile for the physical quantity $U$ over volume, for halos of mass $M$:
    \begin{equation}\label{eq:UM}
      \tilde{U}(M)\equiv4\pi\int_0^\infty dr\,r^2\,U(r|M),
    \end{equation}
    where $U(r|M)$ is the halo profile. $\bU$ is thus the cosmic average value of $U$ weighted by the linear bias of the haloes it populates. This is an interesting quantity to measure, as it can complement other measurements of the unweighted cosmic average $\langle U\rangle$, by providing a handle on the mass dependence of $U$. Note that, although we have used the matter power spectrum in Equation~\ref{eq:pk_hm}, the simplest halo model prediction uses the \emph{linear} matter power spectrum instead. Both agree on sufficiently large scales, although we will explore different options for $P_{mm}(k,z)$ when building robust tomographic reconstruction estimators in Section~\ref{ssec:theory.estimator}.
    
    Consider now the cross-correlation between galaxies at a given mean redshift $\bar{z}_g$, and a projected field $u$, $C^{gu}_\ell$, and the galaxy auto-correlation $C_\ell^{gg}$. As discussed above, on sufficiently large scales, and assuming that the width of the galaxy redshift distribution is much smaller than the timescales over which cosmological quantities and biases vary significantly, we can model these two power spectra as:
    \begin{equation}
      C^{gu}_\ell\simeq b_g\,\bU_{\bar{z}_g}\,T^{gu}_\ell,\hspace{12pt}
      C_\ell^{gg}\simeq b_g^2\,T_\ell^{gg},
    \end{equation}
    where the template angular power spectra $T^{gu}_\ell$ and $T^{gg}_\ell$ are
    \begin{equation}
      T_\ell^{xy}\equiv\int \frac{d\chi}{\chi^2}\,q_x(\chi)\,q_y(\chi)\,P_{mm}\left(\frac{\ell+1/2}{\chi},z(\chi)\right).
    \end{equation}
    Since these templates depend only on the known radial kernels, and on cosmological parameters, the amplitudes of $C_\ell^{gu}$ and $C_\ell^{gg}$ can be used, in combination, to measure $\bU$ at the redshift of the galaxies, in addition to the galaxy bias $b_g$.

    Although we have defined $\bU$ as a bias-weighted average in Equation~\ref{eq:bU}, as predicted by the halo model, this quantity can be more generally interpreted as the relative clustering amplitude of the quantity $U$ with respect to the matter overdensity. Quantitatively, this could be defined as in e.g. \cite{Young2021,Chen2023}:
    \begin{equation}\label{eq:lsa}
      \langle bU\rangle =\lim_{k\rightarrow0}\frac{P_{Um}(k,z)}{P_{mm}(k,z)}.
    \end{equation}
    We will refer to this model as ``large scale approximation'' (LSA), hereafter.
    
    Ultimately, we need to be able to connect the measured amplitude parameter $\bU$ with the clustering of both $U$ and $\delta_m$, with as little dependence as possible on the unknown clustering of the galaxies used to make the measurement.


  \subsection{Tomographic reconstruction estimators}\label{ssec:theory.estimator}
    Consider now the problem of estimating $\bU$ from real measurements of the galaxy auto-correlation and its cross-correlation with the relevant LSS tracer. For simplicity, we will address this problem directly at the level of the 3D power spectra $P_{xy}(k)$, since these measurements are straightforward to make in simulations, and the radial projection onto $C_\ell$s does not introduce any additional relevant complications (assuming that the galaxy redshift distribution is compact and well known).

    Let $\hat{P}_{gg}(k)$ and $\hat{P}_{gU}(k)$ be measurements of the the auto- and cross-correlations, made at a discrete set of wavenumbers $k$. A simple ansatz to model these measurements would be
    \begin{equation}\label{eq:lin_order_bg}
      \hat{P}_{gg}(k)=b_g^2\,\tilde{P}_{mm}(k),\hspace{12pt}
      \hat{P}_{gU}(k)=b_g\langle bU\rangle\,\tilde{P}_{mm}(k),
    \end{equation}
    where $\tilde{P}_{mm}(k)$ is a suitable fixed template for the matter power spectrum. The values of $b_g$ and $\bU$ can then be determined by fitting the measured power spectra to this model.
  
    This model can be improved by adding a constant term in $\hat{P}_{gg}(k)$ to account for the presence of shot noise in the galaxy auto-correlation. For sufficiently dense samples, this term would only be relevant at large values of $k$, but in general shot noise can impact all scales. At the same time, we may include a similar term for $\hat{P}_{gU}(k)$. While in principle shot noise would not affect these cross-correlations, at intermediate scales this term would be able to absorb the small-scale contribution to $\hat{P}_{gU}(k)$ from pairs of points within the same dark-matter halo (i.e. the ``1-halo'' contribution, in the language of the halo model), as well as any other sources of uncorrelated noise in either spectrum.
    The extended model would then read \citep{2206.15394}:
    \begin{align}\label{eq:linPgg}
      &\hat{P}_{gg}(k)=b_g^2\,\tilde{P}_{mm}(k)+N_{gg},\\ \label{eq:linPgU}
      &\hat{P}_{gU}(k)=b_g\langle bU\rangle\,\tilde{P}_{mm}(k)+N_{gU},
    \end{align}
    where $N_{gg}$ and $N_{gu}$ are now two additional free parameters to be included in the fit. We will refer to this model as ``linear model'' or $n_{\text{max}}=0$ (in reference of the more general model explained below).

    Finally, we may further enhance the model by adding terms that could absorb the impact of non-linear biasing in both $\delta_g$ and $U$. Assuming these deviations to be slow-varying in $k$ space, this could be achieved through a polynominal expansion in $k$. Here we will consider adding terms of the form $\propto k^{2n}\tilde{P}_{mm}(k)$ which, for values of $n\geq1$, lead to contributions that are relevant at increasingly small scales, where we expect deviations from linear bias to become more relevant. The most general model we will use here to estimate $\bU$ is thus
    \begin{align}\label{eq:P_gg}
      &\hat{P}_{gg}(k)=b_g^2\,\tilde{P}_{mm}(k)+\sum_{n=1}^{n_{\rm max}}A_{gg}^n\,k^{2n}\,\tilde{P}_{mm}(k)+N_{gg},\\\label{eq:P_Ug}
      &\hat{P}_{gU}(k)=b_g\langle bU\rangle\,\tilde{P}_{mm}(k)+\sum_{n=1}^{n_{\rm max}}A_{gU}^n\,k^{2n}\,\tilde{P}_{mm}(k)+N_{gU},
    \end{align}
    where $A^n_{gg}$ and $A^n_{gU}$ are additional free amplitude parameters. The simplicity of this model allows us to obtain fast analytical estimates for all free parameters, as described in Appendix \ref{app:like}. 

    The final ingredient of this estimator is the template $\tilde{P}_{mm}(k)$ used in lieu of the matter power spectrum. Here we will consider four different options:
    \begin{itemize}
      \item The dark-matter only (DMO) linear matter power spectrum.
      \item The DMO non-linear power spectrum calculated via the \hfit parametrisation of \citep{Takahashi2012}.
      \item The DMO non-linear power spectrum calculated through the \bemu emulator \citep{Angulo2021TheCosmology}.
      \item The matter power spectrum estimated from the same simulation used to measure $\hat{P}_{gg}$ and $\hat{P}_{gU}$. This will include, for example, the impact of baryonic effects \citep{SchallerEtal2024b}.
    \end{itemize}
    The first three options represent viable choices in a realistic data analysis scenario, and differ in the level of accuracy with which they predict the true underlying matter power spectrum. The fourth option is clearly unrealistic, as the true matter power spectrum is not directly observable, but the resulting constraints on $\bU$ represent the best possible scenario, unaffected by inaccuracies in the model of the matter power spectrum.
  
    This defines the most general cross-correlation-based estimator of $\bU$ we will use in this work. We will apply this estimator in Section \ref{sec:res} to different galaxy samples, and quantify the consistency of the resulting $\bU$ measurements. Rather than only comparing the values found for different galaxy samples among themselves, we will compare these with a ``true'' value, independent of any galaxy clustering properties. Instead of predicting $\bU$ according to the halo model, using Equation~\ref{eq:bU}, we will define its reference value as the value found by the estimator defined above using the matter overdensity instead of the overdensity of any galaxy sample (i.e. using $\hat{P}_{mm}(k)$ and $\hat{P}_{mU}(k)$ measured from the simulation instead of $\hat{P}_{gg}(k)$ and $\hat{P}_{gU}(k)$). This will allow us to ascertain if the $\bU$ estimated from galaxies agrees with a universal definition of $\bU$ that only depends on the clustering of matter and of the tracer $U$. 
    
  \subsection{The halo model prediction}\label{ssec:theory.HM}
    Once we have established the agreement between the value of $\bU$ estimated from galaxies, and its reference value estimated from the matter field, we will turn to the question of whether this reference value agrees with the halo model prediction for $\bU$. This theoretical prediction can be calculated by solving the integral in Equation~\ref{eq:bU}, using existing parametrisations for the halo mass function (HMF) and halo bias (e.g. \citep{Tinker08,Tinker10,Despali16,Ondaro21}), as well as a model for the halo mass dependence of $U$. However, the resulting estimate of \ref{eq:bU} would be subject to the inaccuracies in these parametrisations \citep{1104.0949}, as well as differences in the halo mass dependence of hydrodynamical quantities (e.g thermal energy or SFR) in FLAMINGO and in existing models based on first principles or other simulation suites. Instead, we will estimate the halo model prediction by implementing Equation~\ref{eq:bU} directly in the simulation:
    \begin{equation}\label{eq:bU_HM_sim}
      \bU_{\rm HM} = \frac{1}{V_{\rm box}}\sum_h b_h\,\sum_{i\in h}\tilde{U}_i.
    \end{equation}
    Here, the index $h$ runs over all haloes resolved in the simulation, while $i$ runs over all simulation particles identified as belonging to halo $h$. $b_h$ is the linear halo bias for the $h$-th halo, which we calculate in three different ways:
    \begin{itemize}
      \item $b_h$ calculated directly from the simulation. We estimate the value of $b_h$ using the estimator of $\bU$ described in the previous section (Eqs. \ref{eq:P_gg} and \ref{eq:P_Ug}), with $\hat{P}_{mm}(k)$ instead of $\hat{P}_{gg}(k)$, and the matter-halo cross-spectrum, $\hat{P}_{mh}(k|M_h)$, instead of $\hat{P}_{gU}(k)$. See Section~\ref{ssec:bh} for a full description.
      \item $b_h$ calculated from \cite{Tinker10}.
      \item $b_h$ calculated using the Peak-Background Split (PBS) framework with the HMF in \cite{Despali16}
      \item $b_h$ calculated using PBS and the HMF in \cite{Ondaro21}
    \end{itemize}
    Finally, $\tilde{U}_i$ is the value of the $U$ field integrated over the effective volume of the $i$-th particle. To estimate its value for each halo, $\sum_{i\in h}\tilde{U}_i$, we proceed as follows:
    \begin{itemize}
      \item For the case $U=\rsfr$, $\tilde{U}_i$ is simply the total SFR assigned to the particle. Thus, we use the total SFR for each halo in the FLAMINGO halo catalogue. This is calculated by combining all particles assigned to the halo according to a spherical overdensity mass definition with an overdensity $\Delta=200$ times larger than the critical density.
      \item For the case of $U=P_e$, we calculate $\sum_{i\in h}\tilde{U}_i$ from the values of the Compton-$y$ parameter assigned to each halo in the FLAMINGO catalogues. As described in \cite{McCarthy2017}, the volume-integrated integrated pressure in this case is given by
      \begin{equation}
        \tilde{P}_{e,h}=y_h\,d^2_A(z)\,\frac{m_ec^2}{\sigma_T},
      \end{equation}
      where $d_A$ is the angular distance to the redshift of the corresponding snapshot, and $y_h$ is the Compton-$y$ parameter of the halo, calculated by summing over all particles assigned to it. Unlike in the case of SFR, the tSZ signal receives significant contribution from gas at distances beyond the virial radius of the halo \citep{McCarthy:2013qva}. To ensure that this contribution is included in our calculation, in this case we include all particles within a sphere of radius $5\times R_{500c}$, where $R_{500c}$ is the spherical overdensity radius of the halo with an overdensity of $\Delta=500$ times the critical density. Note that this may lead to double-counting of particles lying within the spheres of more than one halo, as discussed in Section \ref{sssec:res.val_bar.halomod}.
    \end{itemize}


\section{Simulations}\label{sec:sims}
  We employ the state-of-the-art \flamingo simulations \citep{SchayeEtal2023,KugelEtal2023}, one of the largest sets of gravity-only (GrO) and hydrodynamical simulations with cosmological volumes ever produced. They are calibrated such that the stellar and AGN feedback models reproduce the observed $z=0$ galaxy stellar mass function and cluster gas fractions, shown to reproduce cluster scaling relations and thermodynamic profiles \citep{Braspenning:2023bmq}. The simulations were run with the SWIFT code \citep{SchallerEtal2024}, using the SPHENIX smoothed-particle hydrodynamic scheme \citep{BorrowEtal2022}.
  
  We focus on the fiducial simulation of the set. This has a cosmology, dubbed D3A, based on the results of the DES-Y3 3x2point analysis, including all external constraints \citep{AbbottEtal2022}. That is, it assumes a flat geometry, a total matter density parameter $\Omega_{\rm m} = 0.306$, a baryon density parameter $\Omega_{\rm b} = 0.0486$, and a neutrino component with total mass $M_\nu = 0.06 ~ \mathrm{eV}$. The power law index of primordial scalar perturbations is $n_{\rm s} = 0.967$ and the amplitude of the primordial power spectrum on a pivot scale $k_{\rm pivot} = 0.05 ~ \mathrm{Mpc}^{-1}$ is $A_{\rm s} = 2.099 \times 10^{-9}$. This corresponds to a standard deviation of linear perturbations (smoothed on spheres of radius $8 ~ \hMpc$) at $z=0$ of $\sigma_{8, \mathrm{tot}} = 0.807$ when considering perturbations in total matter or $\sigma_{8, \mathrm{cb}} = 0.811$ when considering only cold matter perturbations (CDM + baryons). The Hubble parameter at redshift zero is $H_0 = 68.1 ~ \mathrm{km} ~ \mathrm{s}^{-1} ~ \mathrm{Mpc}^{-1}$. The fiducial simulation we employ simultaneously evolves $1800^3$ dark-matter particles, an equal number of initial gas particles, and $1000^3$ neutrino particles in a comoving box of side $L_{\rm box}=1000 ~ \mathrm{Mpc}$. The mass of each dark-matter particle is $M_{\rm DM} = 5.65 \times 10^9 ~ \mathrm{M}_\odot$, and the initial mass of the gas particle is $M_{\rm gas} = 1.07 \times 10^9 ~ \mathrm{M}_\odot$. The particles are initialised at $z=31$ using 3-fluid third-order Lagrangian perturbation theory. Cosmic variance is partially suppressed by enforcing a fixed amplitude initial power spectrum (matching the ensemble average) on scales $(k L_{\rm box})^2 < 1025$. The simulations have been post processed to identify dark matter haloes, their substructures, and galaxies with their respective properties. We use haloes identified with an updated version \citep[][]{Forouhar-MorenoEtal2025} of the Hierarchical Bound Tracing algorithm, \texttt{HBT} \citep{HanEtal2018}. Galaxy properties are computed with a \flamingo-specific tool called SOAP, Spherical Overdensity and Aperture processor \citep[see][for details]{SchayeEtal2023}. Specifically, we use galaxy properties computed by SOAP in apertures of 50 kpc, excluding unbound particles.

  The different galaxy-types used in our analysis have been selected by a cumulative-bin in stellar-mass, $M_{*}$. We sort galaxies in the simulation in decreasing order of $M_{*}$, and select the first $N$ galaxies, with $N=nV_{\text{box}}$. We create three bins with number-densities\footnote{Note that, confusingly, we use ``$h$-inverse'' units to define the galaxy samples used here, as opposed to ``non-$h$'' units used elsewhere in this work. This is to match the same galaxy samples used in \cite{Zennaro25}.} $\ngall$. Note that we used the fiducial simulation with $L_{\rm box}=1000 ~ \mathrm{Mpc}$, instead of the larger-box simulation, with $L_{\rm box}=2800\,{\rm Mpc}$, in order to facilitate a direct comparison with the results presented in \cite{Zennaro25}, which used the same galaxy samples used here. We have tested that the results presented here are not affected by this choice, by comparing the PS of the two simulations at the scales of interest.


\subsection{$P(k)$ estimation}\label{ssec:Pk}
  We calculate new power spectra, independent from \flamingo, for all tracers: simulation particles, halos and galaxies. For all of them we use an FFT grid with $n_g = 512^3$ cubic cells covering the whole simulated volume, and a Triangular Shape Cloud (TSC) mass assignment. This corresponds to a Nyquist frequency of $k_{N} = 1.61 ~ \mathrm{Mpc}^{-1}$. Although this corresponds to a moderate spatial resolution, it is sufficient for the tomography estimators we will explore here, which employ relatively large scales. We use a logarithmic spacing from the fundamental mode to $k_N$, sampling $n_g/4$ bins. The final product is corrected for aliasing following \citet{Angulo2007,Sato2011,Takahashi2012}.

  To construct the U maps from simulated particles, that will then be cross-correlated with $\delta_g$ to create $\hat{P}_{gU}(k)$, we proceed as follows:
  \begin{itemize}
    \item $U = \delta_{m}$, we just add the masses of all particle types in the simulation (dark matter, gas, stars and neutrinos) at each grid point, and divide by the mean mass per cell.
    \item $U = \rho_{\text{SFR}}$: in addition to the TSC weights, each gas particle is weighted by its SFR. The total SFR in each cell is then divided by the comoving cell volume.
    \item $U = P_e$: the comoving thermal pressure is calculated by adding the total thermal energy of all particles in each cell and dividing by its comoving volume. As described in \cite{2005.00009}, this is achieved by weighting each particle by the quantity
    \begin{equation}
      \tilde{P}_{e,i} = k_Bn_{e,i}T_{e,i}\frac{V_i}{V_{\text{cell}}},
    \end{equation}
    where $k_{B}$ is the Boltzmann constant, $n_{e,i}$ is the electron number density for the $i$-th gas particle, $T_{e,i}$ is its electron temperature, and $V_i=m_i/\rho_i$ is the SPH volume element of the particle. The comoving cell volume is simply $V_{\text{cell}}=(L_{\text{box}}/n_g)^3$.
  \end{itemize}

\subsection{$b_{h}$ estimation}\label{ssec:bh}
      \begin{figure*}   
        \begin{subfigure}[b]{\textwidth}
          \centering
          \includegraphics[width=0.475\linewidth]{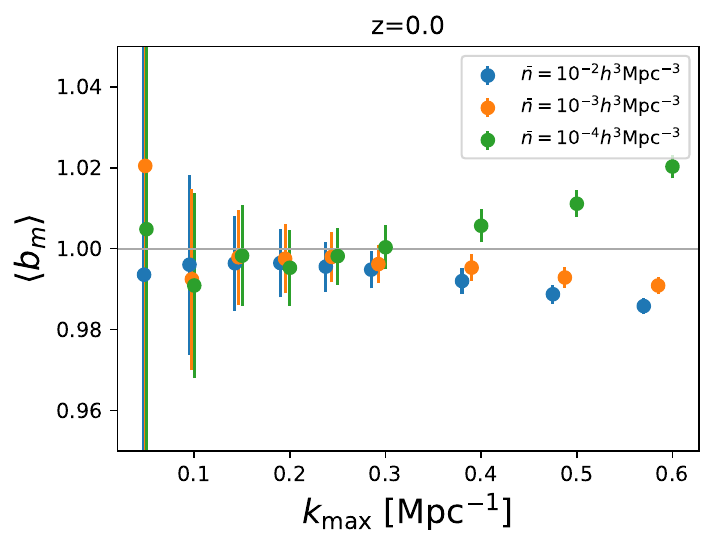}
          \hfill
          \centering
          \includegraphics[width=0.475\linewidth]{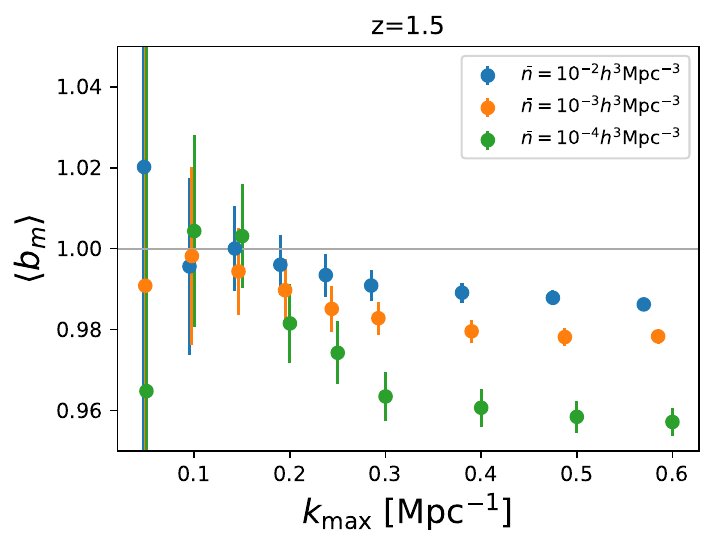}  
          \subcaption[]{Linear bias model, corresponding to Equations~\ref{eq:P_gg} and \ref{eq:P_Ug} with $n_{\text{max}}=0$.}
        \end{subfigure}
        \begin{subfigure}[b]{\textwidth}
          \centering
          \includegraphics[width=0.475\linewidth]{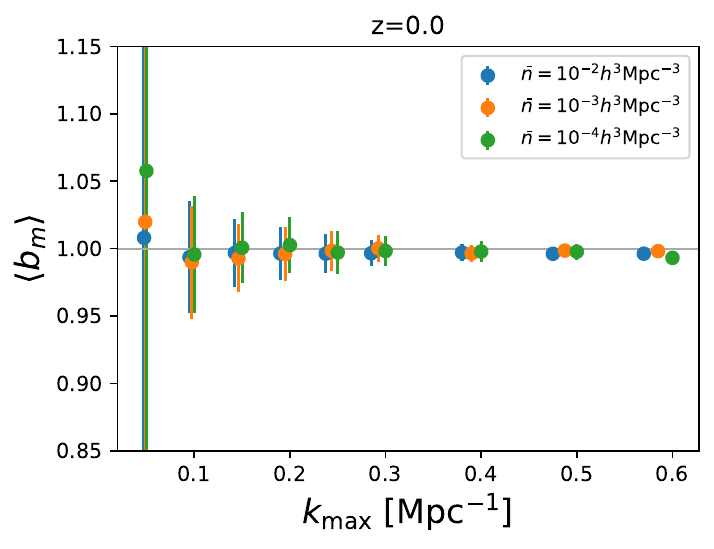}
          \hfill
          \centering
          \includegraphics[width=0.475\linewidth]{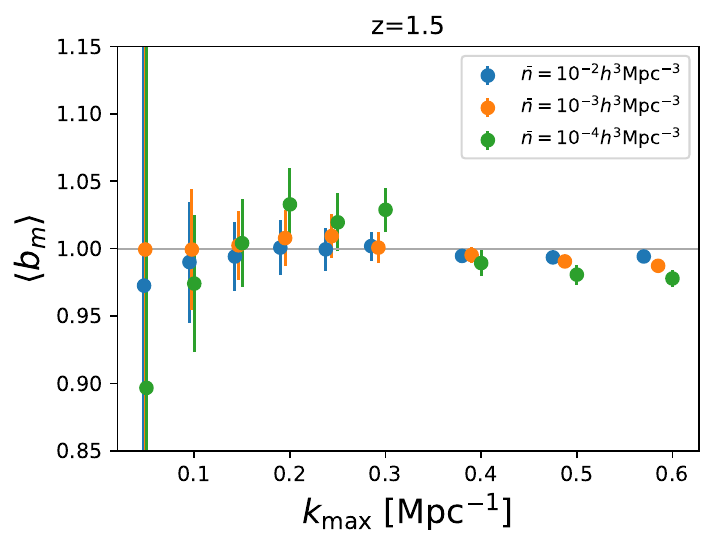}  
          \subcaption[]{Quadratic model, corresponding to Equations~\ref{eq:P_gg} and \ref{eq:P_Ug} with $n_{\text{max}}=1$.}
        \end{subfigure}
        \caption{Deviation from theoretical prediction in the value of $\bM$ as a function of $k_{\text{max}}$, and for two different redshifts ($z=0, 1.5$ in the left and right panels, respectively). Results are shown for three stellar mass-selected samples, with number densities $\ngall$ (blue, orange, and green points, respectively). The top and bottom panels show the results for two different choices of $n_{\rm max}$, which defines the bias model used to fit the data according to Equations \ref{eq:P_gg} and \ref{eq:P_Ug}.}\label{fig:bm}
      \end{figure*}
  In order to calculate the halo model prediction $\langle bU \rangle_{\text{HM}}$ from Equation~\ref{eq:bU_HM_sim}, we need to estimate the halo bias $b_h$ as a function of mass. Although we will test the performance of parametrisations present in the literature \citep{Tinker10, Despali16, Ondaro21}, we also infer its value directly from the simulation. This will ensure that any differences between $\bU_{\rm HM}$ and the value of $\bU$ estimated from the data are not due to inaccuracies in the halo bias parametrisations, or to completeness issues for low mass haloes in FLAMINGO, rather than a genuine failure of the halo model. We measure $b_h(M,z)$ by dividing the FLAMINGO halo catalogue in 5 different mass bins at each snapshot. We bin haloes according to their spherical overdensity mass, with an overdensity parameter $\Delta=200$, defined with respect to the critical density. We use adjacent logarithmic mass bins with edges:
  \begin{equation}
    \log_{10}(M_{200c}/M_\odot)=[8,\,11,\,12,\,13,\,14,\,17].
  \end{equation}

  In each mass bin, we construct maps of the halo overdensity and estimate its cross-correlation with the matter overdensity field $\hat{P}_{mh}(k,z|M)$. We then use this in combination with a measurement of the matter power spectrum $\hat{P}_{mm}(k,z)$ to estimate the halo bias $b_h$ using the same fiducial prescription we will use in Section \ref{sec:res} to estimate $\bU$. The prescription is based on the model in Equations~\ref{eq:P_gg} and \ref{eq:P_Ug}, with $n_{\rm max}=1$, setting $b_g\rightarrow1$ and $\bU\rightarrow b_h$, and using scales up to a maximum wavenumber $k_{\text{max}}=0.3\text{Mpc}^{-1}$.

  Once $b_h$ has been estimated in each mass bin, we build a linear interpolating function associating the measured values of $b_h$ with the median mass in each bin. We then assign a value of $b_h$ to each halo in the simulation by evaluating this function at the corresponding halo mass $M_{200c}$. Note that the first mass bin above is highly incomplete. Nevertheless, the method used here to measure $b_h$ accounts for the impact of this inconsistency in the effective halo bias at low masses.


\section{Results}\label{sec:res}
  In this section we will first study the robustness of the procedure described in Section \ref{ssec:theory.estimator} to reconstruct $\bU$ from galaxy cross-correlations, and select a fiducial version of this procedure simultaneously optimising the precision and robustness of the recovered $\bU$. In doing this, we will also evaluate our ability to account for the dependence of the reconstructed values on cosmological parameters. We will then apply our chosen prescription to the recovery of $\bPe$ and $\brsfr$ from FLAMINGO, verifying that consistent values can be obtained with different galaxy samples, and compare the result with existing measurements from real data. Finally, we will turn to the interpretability of the recovered $\bU$ in terms of the halo model.

  \subsection{Validation of the model}\label{ssec:res.val_bm}
    \subsubsection{Bias modelling and scale cuts}\label{sssec:res.val_bm.kmax}
      Our first task is to determine the range of validity of the bias model described by Equations~\ref{eq:P_gg} and \ref{eq:P_Ug} and used to extract measurements of $\bU$, as a function of galaxy type, $n_{\rm max}$, and of the smallest scale used $k_{\rm max}$. To do so, we will test the model against a known solution, by setting $U=\delta_m$, the matter overdensity. Label the corresponding bias-weighted quantity in this case $\bM$, an unbiased estimator should recover the value $\bM=1$. As a power spectrum template, $\tilde{P}_{mm}(k)$, we used the true matter power spectrum of the hydrodynamical simulation being analysed, in order to factor out uncertainties related with this choice, which we will study in detail in the next section.

      Figure~\ref{fig:bm} shows the value of $\bU$ estimated from our three different galaxy samples (blue, orange, and green points for the samples with number densities $\ngall$, respectively). Results are shown as a function of $k_{\rm max}$, for the linear bias model with $n_{\rm max}=0$ (top panels), and for the quadratic model ($n_{\rm max}=1$, bottom panels). Results are shown for the snapshots at $z=0$ and $z=1.5$ (left and right, respectively). We find that, as expected, the quadratic model is significantly more stable than the linear model, particularly at low redshifts. We find that the latter is able to obtain unbiased results across different samples up to scales $k_{\text{max}}>0.3-0.5\text{Mpc}^{-1}$, whereas the linear model breaks down dramatically at $k_{\text{max}}=0.15-0.25\text{Mpc}^{-1}$. This, as we will see in Section \ref{ssec:res.val_bar} is also comparable to the performance achieved when recovering $\brsfr$ and $\bPe$.

\begin{figure}
        \centering
        \includegraphics[width=0.49\textwidth]{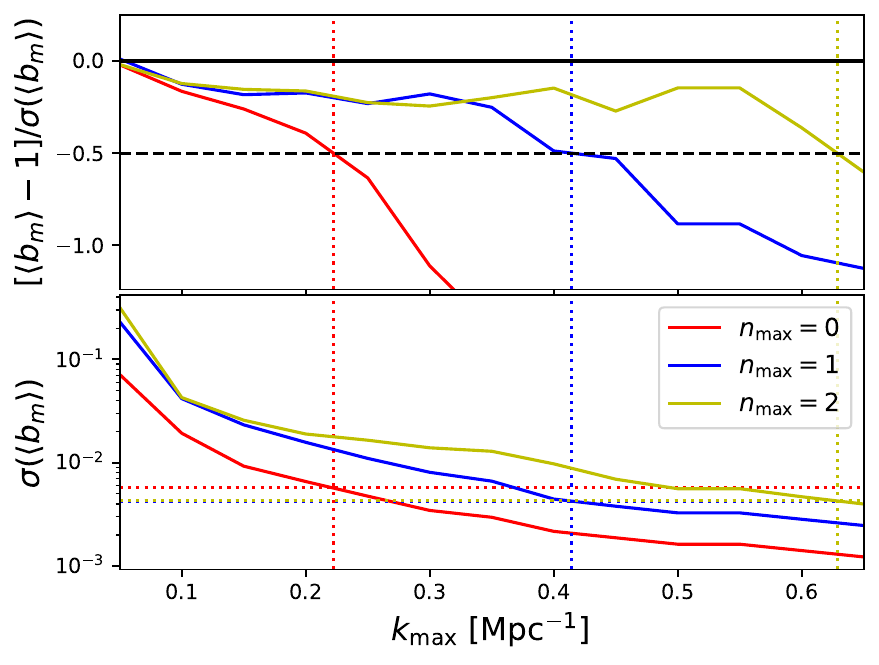}
        \caption{\emph{Top panel:} deviation from the theoretical prediction in the value of $\bM$, relative to the statistical uncertainties in this quantity, as a function of $k_{\text{max}}$. Results are shown for bias models with $n_{\rm max}=0$, 1, and 2 (red, blue, and yellow contours, respectively). The dotted lines mark the scales at which the bias in $\bM$ exceed $0.5\sigma$. \emph{Bottom panel:} uncertainty on the recovered $\bM$ as a function of $k_{\rm max}$ for the same bias models, with the dotted lines marking the smallest uncertainty achieved by each model within the range of $k_{\rm max}$ for which it remains unbiased. All results are shown for the galaxy sample with density $\ngal{2}$ at an intermediate redshift $z=0.8$.}
        \label{fig:bm_k4}
      \end{figure}

      Increasing $n_{\rm max}$ leads to increased stability in the recovered value of $\bU$, and robustness against deviations from a purely linear bias relation in both $\delta_g$ and $U$. However, this comes at the cost of larger statistical uncertainties. It is therefore interesting to study if an optimal value of $n_{\rm max}$ can be found that minimises the error in the resulting measurement of $\bU$ within the range of scales for which the estimator remains sufficiently unbiased. Figure~\ref{fig:bm_k4} shows the results of inferring $\bM$ as a function of $k_{\rm max}$ for different values of $n_{\rm max}\in\{0,1,2\}$. The results shown were obtained for the galaxy sample with $\ngal{2}$ at an intermediate redshift $z=0.8$, but the same qualitative results are obtained for other galaxy-redshift combinations. The top panel shows the difference in the recovered value of $\bM$ with respect to the theoretical expectation of $\bM=1$ as a fraction of the statistical uncertainties. the horizontal dashed line marks the points at which the estimator recovers a bias of $\sim0.5\sigma$\footnote{Note that this benchmark (a bias in $\bM$ equal to half the statistical uncertainties) is relatively arbitrary, as $\sigma(\bM)$ is largely determined by the size of the simulation box. We use it here only to highlight the trade-off between precision and accuracy when selecting a value of $n_{\rm max}$. Elsewhere in this work, we will use the fractional bias on $\bU$ as a metric to quantify the performance of a given estimator.}, with the vertical dotted lines signalling the scale cut $k_{\rm max}$ at which this happens. The bottom panel then shows the error with which $\bM$ is recovered, with the dotted lines showing the error achieved at the corresponding $k_{\rm max}$. We can see that, while the uncertainties for $n_{\rm max}=0$ are consistently lower than for $n_{\rm max}=1$ or 2 at any given $k_{\rm max}$, the fact that $n_{\rm max}=1$ can recover unbiased results while pushing to smaller scales allows it to outperform $n_{\rm max}=0$. In turn, while $n_{\rm max}=2$ allows us to use even smaller scales while remaining unbiased, it does not outperform $n_{\rm max}=1$. Given this, we choose $n_{\rm max}=1$ as the fiducial bias model defining our estimator for $\bU$, balancing precision and robustness against non-linear biasing. With this choice, as we will show also in the cases of $\bPe$ and $\brsfr$, we will in general be able to obtained largely unbiased constraints using scales up to $k_{\rm max}=0.3\,{\rm Mpc}^{-1}$. 

    \subsubsection{Power spectrum template}\label{sssec:res.val_bm.pk}
\begin{figure}
      \begin{subfigure}[b]{0.5\textwidth}
        \centering
        \includegraphics[width=0.95\linewidth]{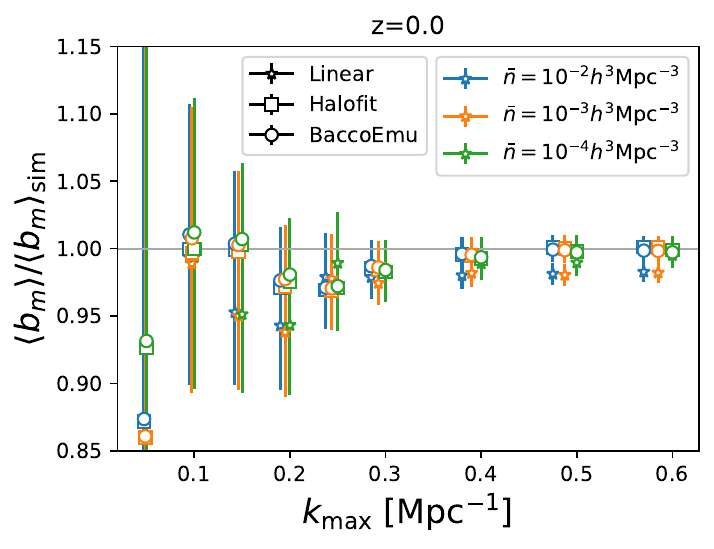}
      \hfill
        \centering
        \includegraphics[width=0.95\linewidth]{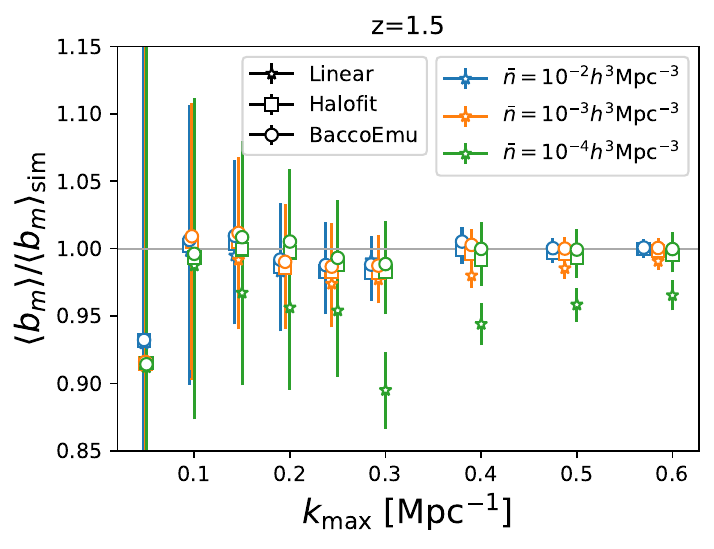}  
      \end{subfigure}
      \caption{Estimation of $\bM$ in the quadratic model (i.e. $n_{\rm max}=1$)    with $P(k)$ from linear-theory, and non-linear \hfit and \bemu results (star, squared and circle, respectively), as a function of $k_{\text{max}}$ for two different redshifts ($z=0, 1.5$). We study galaxies selected in order of decreasing stellar mass, with the different colours (blue, orange, and green) corresponding to different number densities, namely $\ngall$. }
      \label{fig:Pkcomparison_bm}
      \end{figure}

      Another variable defining our estimator of $\bU$ is the choice of power spectrum template $\tilde{P}_{mm}(k)$ used to fit the data. The previous analysis has been done using the spectrum measured directly from the simulation $\tilde{P}_{mm}^{\rm sim}(k)$. This is obviously not available when applying this estimator to real data, and a model predicting $\tilde{P}_{mm}(k)$ must be used. In Figure~\ref{fig:Pkcomparison_bm} we test the robustness of our results with respect to the choice of template, exploring the three other templates described in Section \ref{ssec:theory.estimator}: the linear power spectrum, the \hfit prediction for the non-linear matter power spectrum, and the result of the \bemu emulator for the same quantity. The results shown were obtained for our fiducial bias model with $n_{\rm max}=1$, and we present them for the $z=0$ and $z=1.5$ snapshots (top and bottom panels). 
      
      We find that both \hfit and \bemu are able to obtain unbiased constraints on $\bM$ for all galaxy samples up to relatively small scales (certainly within the range $k_{\rm max}\leq0.3$). Even the linear matter power spectrum recovers relatively accurate constraints, although the results are substantially less stable, particularly for the most massive galaxy sample explored ($\ngal{4}$) at high redshifts. Thus we find that a relatively naive prediction for the matter power spectrum, such as \hfit, is sufficient to recover robust constraints on $\bU$, with the free parameters of our bias model able to absorb the small-scale inaccuracies of this prediction. It is also interesting to note that, although the simulated data contains baryonic effects, here we use the power spectrum of only the cold matter components (dark matter and baryons) in the absence of baryonic feedback. The bias model is again flexible enough to absorb these effects and yield an unbiased estimate of $\bU$.

    \subsubsection{Dependence on the fiducial cosmology}\label{sssec:res.val_bm.cosmo}
      \begin{table} \centering
      \caption{Cosmological parameters used to quantify the cosmology dependence of the tomographic estimator developed here. The fiducial cosmology corresponds to that of the \flamingo simulation analysed here.}
      \label{tab:cosmo}
      \begin{tabular}{l|lll}
      Name            & $\sigma_8$ & $\Omega_m$ & $n_s$ \\ \hline
      Fiducial        & 0.8069     & 0.3060     & 0.967 \\
      Low $\sigma_8$  & 0.7499     & 0.3060     & 0.967 \\
      Low $\Omega_m$  & 0.8071     & 0.27       & 0.967 \\
      High $\Omega_m$ & 0.8071     & 0.344      & 0.967 \\
      Low $n_s$       & 0.8069     & 0.3060     & 1.0  
      \end{tabular}
      \end{table}

      \begin{figure}
        \centering
        \includegraphics[width=0.95\linewidth]{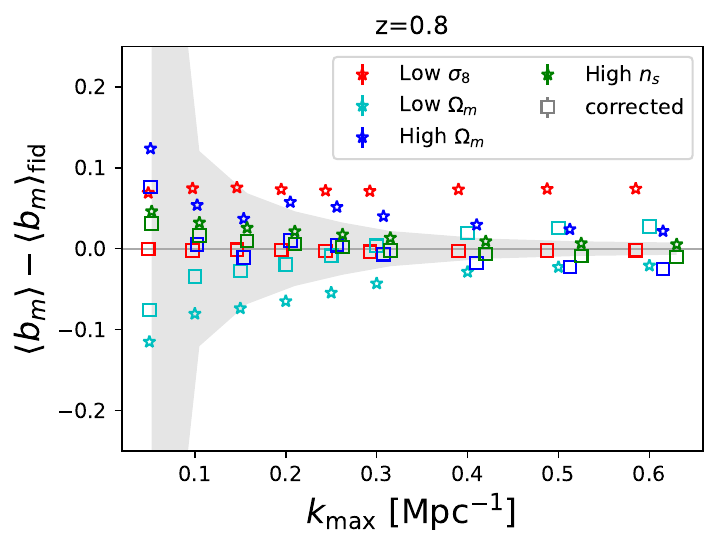}
        \caption{Comparison between the inferred values of $\bM$ obtained when using a $\tilde{P}_{mm}(k)$ with a different cosmology from the fiducial (that of the galaxy measurements). The analysis uses $n_{\text{max}}=1$, $\tilde{P}_{mm}(k)$ from BaccoEmu, and only the matter distribution (baseline fit) is shown. The corrected values are obtained by applying Equation~\ref{eq:dep_cosmo} to the starred values, showing the validity of the model up to the limit scales ($k_{\text{max}}<0.3\text{Mpc}^{-1})$. The shaded area corresponds to the uncertainty in $\bM_{\text{fid}}$}\label{fig:cosmo}
      \end{figure}
      Although we have shown that the estimator is relatively robust to the choice of power spectrum template, this has been under the assumption that this template was created assuming a cosmological model that matches that of the data. This is not possible in practice, and it is therefore important to characterise the dependence of the estimated $\bU$ on the true cosmological parameters, which in general differ from those used to generate the $P(k)$ template.

      Let us consider a simple scenario in which the only parameter mismatch is in the value of the amplitude of matter fluctuations, $\sigma_8$, such that, on the large scales we use to measure $\bU$, $\tilde{P}_{mm}(k)=(\tilde{\sigma}_8/\sigma_8)^2P_{mm}(k)$, where $P_{mm}(k)$ is the true underlying power spectrum, and $\tilde{\sigma}_8$ is the value of $\sigma_8$ assumed to generate $\tilde{P}_{mm}(k)$. On large scales we can equate the $gg$ and $gU$ spectra to
      \begin{align}
        &\hat{P}_{gg}(k)=\hat{b}_g^2\tilde{P}_{mm}(k)=b_g^2P_{mm(k)}\\
        &\hat{P}_{gU}(k)=\hat{b}_g\,\hat{\bU}\tilde{P}_{mm}(k)=b_g\,\bU P_{mm(k)},
      \end{align}
      where hatted quantities (e.g. $\hat{P}$) are those estimated from the data. Solving for $\hat{\bU}$, we find that its dependence on the true value of $\sigma_8$ is, unsurprisingly
      \begin{equation}\label{eq:dep_s8}
        \hat{\bU}=\frac{\sigma_8}{\tilde{\sigma}_8}\bU.
      \end{equation}
      In practice, a general shift in cosmological parameters will result in changes in both the amplitude and the shape of the power spectrum template. We can, however, assume that the additional terms in our generic bias model ($N_{xy}$ and $A_{xy}^n$ in Equation~\ref{eq:P_Ug}) will be able to absorb most of the change in the power spectrum shape while recovering the small-scale behaviour of the measured spectra, while $b_g$ and $\bU$ capture the overall amplitude of the power spectrum on large-scales. With this intuition in mind we will use the following ansatz to capture the cosmology dependence of the estimated value of $\bU$, by generalising Equation~\ref{eq:dep_s8} to
      \begin{equation}\label{eq:dep_cosmo}
        \hat{\bU}=\frac{\sigma(k_L)}{\tilde{\sigma}(k_L)}\bU,
      \end{equation}
      where $\sigma(k_L)$ is the variance of the matter overdensity on wavenumbers smaller than $k_L$:
      \begin{equation}
        \sigma(k_L)=\left[\frac{1}{2\pi^2}\int_0^{k_L}dk\,k^2\,P_{mm}(k)\right]^{1/2},
      \end{equation}
      and $\tilde{\sigma}(k_L)$ is the same quantity calculated for the power spectrum template $\tilde{P}_{mm}(k)$ used in the estimator. We will use a default value of $k_L=0.07\,h\,{\rm Mpc}^{-1}$.

      To validate this ansatz, we reanalyse the simulation data used to infer $\bM$ using matter power spectrum templates generated using cosmologies that differ from the fiducial model (i.e. the fiducial \flamingo cosmology). In particular, we tested cosmologies with a lower value of $\sigma_8$, lower and higher values of $\Omega_m$, and a higher value of the scalar spectral index $n_s$. The specific values used are listed in Table~\ref{tab:cosmo}. Note that these parameter variations are substantially larger than the current uncertainties on these parameters from CMB data. The results of this analysis are shown in Figure~\ref{fig:cosmo}. The figure shows the values of $\bM$, for the baseline marker (the matter density), recovered using power spectrum templates for each cosmology (star markers), as well as the same measurements corrected for the cosmology-dependent factor in Equation~\ref{eq:dep_cosmo} (empty square markers). We can see that the large scatter between different cosmologies is significantly reduced after applying the cosmology-dependent scaling factor, recovering the value of $\bM$ inferred with the unbiased power spectrum template within $1\sigma$ in all cases, within the regime of validity of our estimator ($k_{\rm max}\leq 0.3\,h{\rm Mpc}^{-1}$). Interestingly, the low-$\Omega_m$ cosmology displays the largest disagreement after this correction. It is likely possible to develop a more precise scheme to account for the cosmology dependence of our measurements. However, since the main aim of this paper is to present a scheme for tomographic reconstruction that is robust to the clustering properties of the galaxy samples used, we leave such a study for future work.


\subsection{Tomographic reconstruction of baryonic properties}\label{ssec:res.val_bar}
  Having tested the robustness of the tomographic estimator of $\bU$ in the previous section, we now apply it to the reconstruction of baryonic properties, namely the bias-weighted electron pressure $\bPe$ and SFR density $\brsfr$. The two aims of this section will be:
  \begin{enumerate}
      \item To test the consistency of the estimated $\bU$ across different galaxy samples, and its agreement with the underlying true value (see below). We will do this in Sections \ref{sssec:res.val_bar.SFR} and \ref{sssec:res.val_bar.Pe}.
      \item To study the possibility of interpreting the recovered values of $\bU$ from first principles, using the halo model, and to quantify the limitations of this approach. This will be discussed in Section \ref{sssec:res.val_bar.halomod}.
  \end{enumerate}
  To assess the unbiasedness of $\bU$ estimated from galaxy cross-correlations in Sections \ref{sssec:res.val_bar.SFR} and \ref{sssec:res.val_bar.Pe}, we will consider its true underlying value to be that obtained by applying the same estimator to correlations involving the matter overdensity. While this definition is not provided in terms of fundamental ingredients (describing e.g. the detailed relation between matter, gas pressure, and SFR), the value of $\bU$ thus constructed is a well-defined quantity that can be constructed from either a simulation or a sufficiently accurate model of the matter-matter and matter-$U$ power spectra. Section \ref{sssec:res.val_bar.halomod} will then determine the halo model's accuracy in predicting this quantity.

  Based on the results shown in the previous section, in what follows we will use our tomographic estimator with $n_{\rm max}=1$ (i.e. including scale-dependent biasing terms proportional to $k^2\tilde{P}_{mm}(k)$), employing as template the matter power spectrum measured in \flamingo. For this choice, we determined that the estimator remains robust up to scales $k_{\rm max}\simeq0.3\,{\rm Mpc}^{-1}$ in the case of $\bM$, and we will now study if this is also the case for $\bPe$ and $\brsfr$.
  
  \subsubsection{Estimating $\brsfr$}\label{sssec:res.val_bar.SFR}
    \begin{figure}
        \centering
        \includegraphics[width=0.95\linewidth]{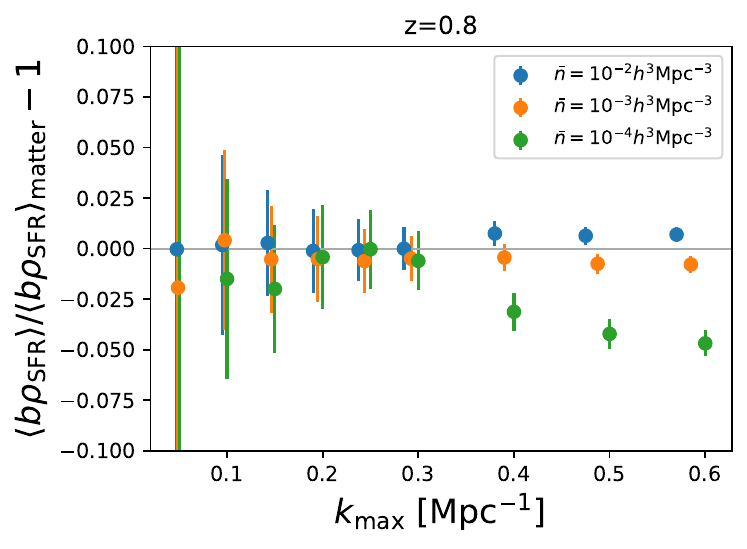}
      \caption{Estimation of $\langle b\rho_{\text{SFR}} \rangle$, using the quadratic model ($n_{\rm max}=1$), as a function of $k_{\text{max}}$ for two different redshifts ($z=0, 1.5$). We study galaxies selected in order of decreasing stellar mass, with the different colours (red, blue, green) corresponding to different number densities, namely $\ngall$. }\label{fig:brhoSFR}
    \end{figure}
    \begin{figure}
      \centering
      \includegraphics[width=\linewidth]{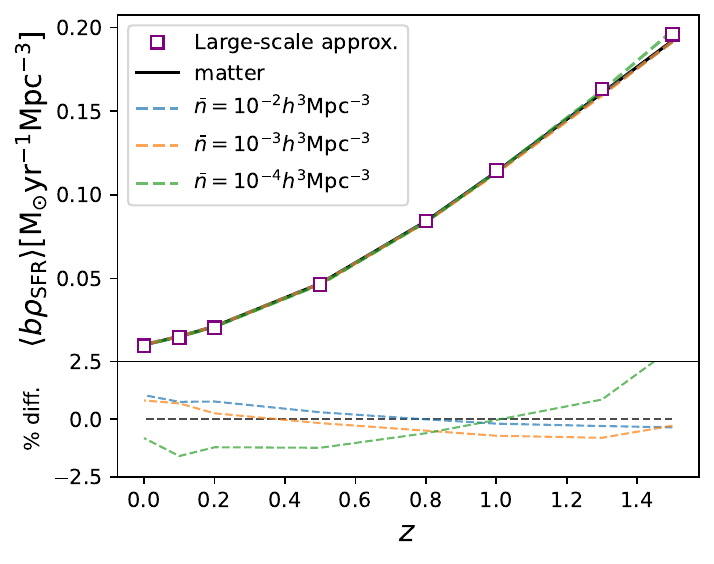}
      \caption{Best-fit results of $\langle b\rho_{\text{SFR}} \rangle$ from the quadratic model ($n_{\rm max}=1$) as a function of redshift, with $P(k,z)=P_{mm,\text{sim}}$, at scales $k_{\text{max}}=0.3\text{Mpc}^{-1}$. We study galaxies selected in order of decreasing stellar mass, with the different colours (blue, orange, and green) corresponding to different number densities, namely $\ngall$. We compare the results obtained from these samples with that found using the matter overdensity. Purple squares correspond to values computed from the large-scale approximation model, using scales $k<0.15\text{Mpc}^{-1}$. The \emph{bottom panel} shows the percent relative difference between the values recovered from the matter field and the different galaxy samples.}\label{fig:brhoSFRcomparison}
    \end{figure}
    We start by applying the tomographic reconstruction estimator to cross-correlations between the galaxy overdensity and 3D maps of the SFR density (as a proxy for CIB cross-correlations). Figure~\ref{fig:brhoSFR} shows the relative difference between the value of $\brsfr$ reconstructed from galaxy cross-correlations and the true value estimated from the cross-correlation with the matter overdensity. Results are presented as a function of the maximum wavenumber included in the analysis $k_{\rm max}$, for our three galaxy samples, with densities $\ngall$, and at an intermediate redshift $z=0.8$. We find that our estimator is able to recover a value of $\brsfr$ that is consistent across galaxy samples and unbiased within $1\sigma$ when using scales $k<k_{\rm max}=0.3\,{\rm Mpc}^{-1}$. This regime of validity is similar to that found in the case of $\bM$, in Section \ref{ssec:res.val_bm}. Furthermore, in this range of scales, we are able to obtain an unbiased estimate of $\brsfr$, accurate at the $\sim1\%$ level, within $1\sigma$ for the statistical uncertainties corresponding to the \flamingo simulation and, more importantly, much smaller than the statistical uncertainties in real data \citep[see Fig. \ref{fig:bU_Sim_vs_HM} and ][]{2206.15394}.

    This result is also consistent across different redshifts. Figure~\ref{fig:brhoSFRcomparison} shows the value of $\brsfr$ found using our fiducial estimator with $k_{\rm max}=0.3\,{\rm Mpc}^{-1}$ for different \flamingo snapshots, spanning the redshift range $z<1.5$. The results are shown for the three different galaxy samples (blue, green, and orange colours), as well as the true value estimated from the cross-correlation with the matter overdensity (in black). The estimates of $\brsfr$ from different galaxy samples agree with each other and with the truth across all redshifts, with percent-level accuracy. The mean relative deviation is $1.2\%$, with the largest deviation, corresponding to $\ngal{4}$ sample at the highest redshift being $3\%$.

    For context, Figure~\ref{fig:brhoSFRcomparison} shows the result of estimating $\brsfr$ using the large-scale approximation of Equation~\ref{eq:lsa}. In this case, we measure $P_{mm}(k)$ and $P_{m\rho_{\rm SFR}}(k)$ from the simulation, take the ratio of both measurements, and estimate $\brsfr_{\rm LSA}$ as the average of this ratio on scales $k<0.15\,{\rm Mpc}^{-1}$. We see that, in this case, the LSA is in fact an excellent approximation to the value of $\brsfr$ calculated through our fiducial estimator. This is not a trivial statement, as scale-dependent and stochastic biasing in the SFR density field could induce a bias in the value of $\brsfr$ inferred through the LSA estimator.


  \subsubsection{Estimating $\bPe$}\label{sssec:res.val_bar.Pe}
    \begin{figure}
      \includegraphics[width=0.95\linewidth]{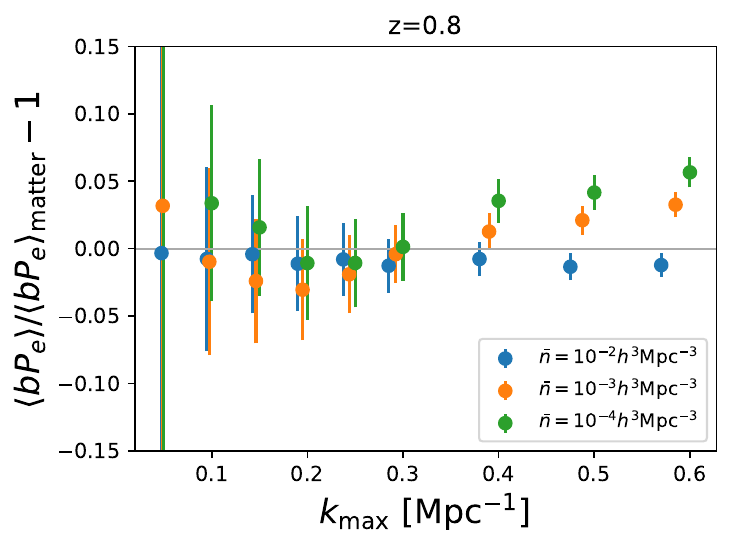}
      \caption{Same as Figure~\ref{fig:brhoSFR}, but for $\langle bP_{e}\rangle$.}\label{fig:bPe}
    \end{figure}

    \begin{figure}
      \centering
      \includegraphics[width=\linewidth]{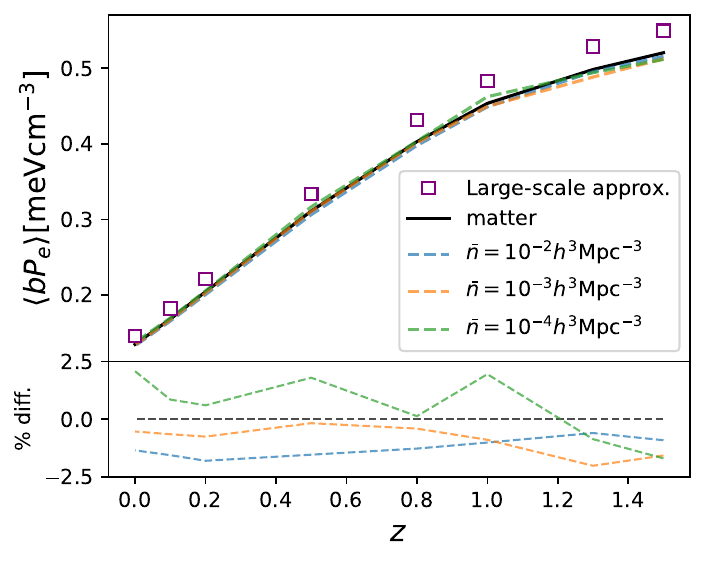}
      \caption{Same as Figure~\ref{fig:brhoSFRcomparison}, but for $\langle bP_{e}\rangle$.}\label{fig:bPecomparison}
    \end{figure}
    We repeat the same procedure on cross-correlations involving maps of the thermal electron pressure, constructed as described in Section \ref{ssec:Pk}, in order to estimate $\bPe$. This is interesting, not only because this is a quantity that can be measured from cross-correlations with maps of the tSZ effect, but because inhomogeneities in the thermal gas pressure and the SFR density have markedly different clustering properties that could affect the performance of the tomographic reconstruction estimator. While star formation efficiency peaks at intermediate halo masses \citep[$\log (M/M_\odot)\sim12$][]{1705.05373,2209.05472,2006.16329}, thermal energy scales strongly with halo mass, and thus thermal pressure maps are dominated by the contribution of a small number of massive haloes \citep{1806.04786}. Both quantities thus probe different cosmic density environments, affecting their their effective scale-dependent and stochastic bias relation.
    
    In Figure~\ref{fig:bPe} we show the relative difference between the values of $\bPe$ recovered from our fiducial estimator for the three different galaxy samples, and the true value recovered using the matter instead of galaxy overdensity. The results are shown as a function of $k_{\rm max}$ for the snapshot at redshift $z=0.8$. As in the case of $\bM$ and $\brsfr$, we find that the estimator is robust against non-linear galaxy biasing up to scales $k>0.3\,{\rm Mpc}^{-1}$, beyond which the estimates from different galaxy samples start to differ significantly. It is interesting to note that the relative statistical uncertainties of the $\bPe$ measurements are noticeably larger than those of $\brsfr$ (see Figure~\ref{fig:brhoSFR}). This may be a consequence of the aforementioned dominant contribution to $P_e$ from sparse massive haloes, raising the effective shot noise contribution to the cross-spectrum measurement.
    
    This result holds across different redshifts, as demonstrated by Figure~\ref{fig:bPecomparison}, which shows the value of $\bPe$ recovered for the different galaxy samples, as well as the true value, for the different redshift snapshots studied here, using scales up to $k_{\rm max}=0.3$. The average relative difference with respect to the truth across all the recovered values is approximately $1.2\%$, with the largest difference being $2\%$, again corresponding to the $\ngal{4}$ sample at the highest redshift probed. As in the previous section, we also show the prediction based on the large-scale approximation, calculated as described above. In this case, however, we find that the LSA significantly overpredicts the value of $\bPe$ at all redshifts. This is not entirely surprising: as mentioned above, the strong influence of massive haloes in the pressure map means that the matter-pressure cross-correlation receives a large 1-halo contribution \citep[see e.g. Fig.~2 of][]{2005.00009}, which acts as an effective shot noise-like term on large scales. Unlike our estimator, the LSA does not account for this component, leading to an over-estimation of the true large-scale amplitude governed by $\bPe$.


  \subsubsection{Theoretical interpretation of tomographic reconstruction}\label{sssec:res.val_bar.halomod}
    \begin{figure}
      \begin{subfigure}{\linewidth}
      \centering
      \includegraphics[width=\linewidth]{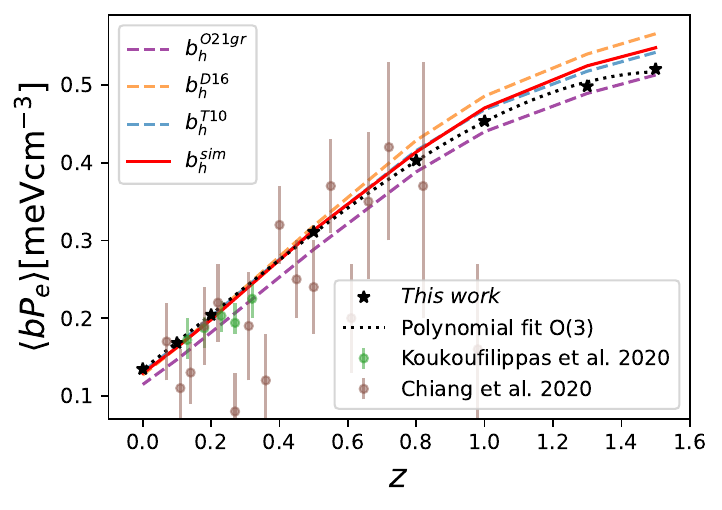}
      \end{subfigure}
      \begin{subfigure}{\linewidth}
      \centering
      \includegraphics[width=\linewidth]{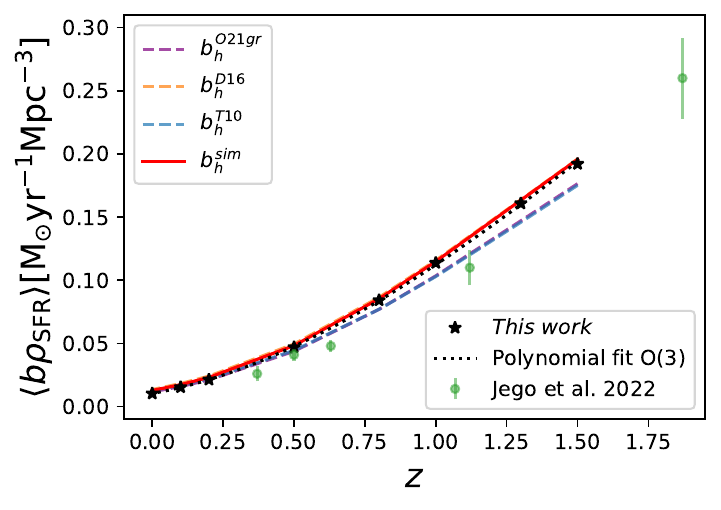}        
      \end{subfigure}
      \caption{Comparison of $\bU = \langle bP_{e} \rangle$ and $\langle b\rho_{\text{SFR}} \rangle$ as a function of redshift for: galaxies (black) and their polynomial linear fit (black dotted line), halo-bias weighted models: \citet{Tinker10} - blue -, \citet{Despali16} - orange -, and \citet{Ondaro21} - purple -, and different real data (see refs in legend).}
      \label{fig:bU_Sim_vs_HM}
    \end{figure}
    So far we have compared our estimator of $\bU$ using galaxy cross correlation against its true value, having defined the latter as ``the result of applying our estimator to correlations involving the matter overdensity''. Although this is a well-defined quantity, requiring only a physical model for the $P_{mm}(k)$ and $P_{mU}(k)$ power spectra, in terms of both cosmology and the astrophysics of the $U$ field, it is interesting to explore whether a more direct theoretical prediction of $\bU$ can be constructed, independent of the specific definition of our estimator.

    This can potentially be achieved via the halo model, which predicts $\bU$ to be given by Equation~\ref{eq:bU}. The halo model prediction depends on three quantities, all of which are, to some extent, uncertain:
    \begin{itemize}
      \item The halo mass function $n(M)$, which depends not only on the chosen halo mass definition, but also on the specific halo finder used to estimate it from simulations \citep{1104.0949,Maleubre2024,Forouhar-MorenoEtal2025}.
      \item The linear halo bias $b_h(M)$, which compounds the same sources of uncertainty as $n(M)$ in addition to potential departures from the predictions of the peak-background split, or the specific method used to estimate it in simulations.
      \item A mass-observable relation describing $\bar{U}(M)$ (defined in Equation~\ref{eq:UM}). In most cases this will involve a model for the astrophysics of $U$, which should be sufficiently accurate or flexible, but will also be affected by uncertainties in the choice of halo definition and its boundaries.
    \end{itemize}
    We would like to separate the potential inaccuracies of the halo model prediction caused by uncertainties in the $n(M)$ and $b_h(M)$ parametrisations, as well as those due to incomplete astrophysical models describing $\bar{U}(M)$, from those arising from non-linear halo biasing and assembly bias. While the former sources of error may be addressed by a more careful design of the models describing these quantities, the latter point to problems in the core assumptions of the vanilla halo model that would require the introduction of new ingredients. To that end, we will aim to use the abundance, linear clustering bias, and mass-observable relations measured directly in \flamingo, instead of choosing any particular parametrisation. We will achieve this by writing Equation~\ref{eq:bU} as a sum over the contribution of all haloes in the simulation, as given in Equation~\ref{eq:bU_HM_sim}, with the halo bias $b_h$ measured directly from the simulation as described in Section \ref{ssec:bh}.

    Figure~\ref{fig:bU_Sim_vs_HM} shows the halo model prediction for $\bPe$ and $\brsfr$ (top and bottom panels, respectively) as a function of redshift, together with the measurements of $\bU$ extracted using our fiducial estimator with $k_{\rm max}=0.3\,{\rm Mpc}^{-1}$ (results are shown for the matter density). In the case of $\brsfr$ we see that the halo model using the halo bias measured directly in \flamingo is able to accurately predict our measurements at all redshifts. The largest relative difference with respect to the measurements is $<2\%$ at the highest redshift. In the case of $\bPe$ we see that the halo model is again precise at low redshifts, but becomes increasingly inaccurate beyond $z\simeq0.6$, consistently overpredicting the measurements.
    
    As we discussed in Section \ref{ssec:theory.HM} (see also \cite{McCarthy2017}), a significant contribution to the pressure map comes from gas in the outskirts of haloes, which motivated us to calculate the total pressure integrated over the halo volume defined as a sphere with radius $R=5\times R_{500}$. However, if the density of haloes that significantly contribute to the pressure map is sufficiently high, this procedure has the risk of double-counting the contribution from simulation particles simultaneously assigned to different haloes within their $5\times R_{500}$ spheres. The total comoving volume occupied by these spheres at $z=1.5$ is approximately three times larger than at $z=0$, significantly increasing the fraction of doubly-counted particles. We verified that using smaller radii to calculate the volume-integrated pressure had a strong effect on the result, particularly at high redshifts. We thus conclude that the definition of halo boundaries may be a major source of uncertainty for the halo model predictions of $\bPe$, particularly at high redshifts.
    
    It is also interesting to place these differences in the context of the theoretical uncertainties associated with the halo model when using existing parametrisations to model its different ingredients. As a way to quantify this, Figure~\ref{fig:bU_Sim_vs_HM} shows the halo model predictions obtained by replacing the halo bias measured directly in \flamingo with various fitting functions. Namely, we use the parametrisation of $b_h$ presented in \cite{Tinker10}, as well as the halo bias derived from the mass function parametrisations of \cite{Despali16} and \cite{Ondaro21}, assuming a relation between both quantities given by the peak-background split \citep{astro-ph/9512127}. We see that using different bias parametrisations leads to differences in the prediction of $\bU$ that are larger than the differences between our fiducial halo model predictions and our measurements. A more detailed discussion of the differences between these halo bias prescriptions is presented in Section \ref{app:bh}. Therefore, the inaccuracy of the halo model in predicting $\bU$ is likely dominated by the theoretical uncertainty in existing parametrisations of the halo model ingredients, rather than in the core assumptions of the halo model itself. 

    Finally, Figure~\ref{fig:bU_Sim_vs_HM} also shows some of the existing measurements of $\brsfr$ from \cite{2209.05472}, and $\bPe$ from \cite{1909.09102} and \cite{2006.14650}. The \flamingo predictions for both of these quantities appear to be in qualitatively good agreement with these measurements. Although a comparison of $\bPe$ measurements against predictions from hydrodynamical simulations was presented in \cite{Chen2023}, this is the first time, to our knowledge, that a similar comparison has been made for $\brsfr$. A more detailed study of the predictions from other simulations in the \flamingo suite, as well as other simulations and theoretical frameworks, against existing measurements could therefore be used to place constraints on both cosmology and the astrophysics of galaxy formation and the intergalactic medium. Once applied to real data, the methodology proposed and validated in this work will ensure that any measurements used in this comparison will be robust against uncertainties in the clustering properties of the galaxies used to extract them.

\section{Conclusions}\label{sec:conc}
  Tomographic reconstruction, the analysis of the cross-correlation between galaxies and projected tracers of other LSS probes, in conjunction with the auto-correlation of these galaxies, can be used to obtain precise measurement of the cosmic average of key physical properties weighted by the halo bias. The resulting measurements can be used to constrain both cosmological parameters and models describing the baryonic components of the cosmic density field. In this paper, we have investigated the possibility of designing a tomographic reconstruction estimator that is robust against the complex clustering properties of the galaxies used in the analysis, as well as some of the challenges that must be tackled when interpreting the resulting measurements.

  We have demonstrated that such an estimator can in fact be constructed, able to recover unbiased estimates of $\bU$ with percent-level accuracy. The method consists of modelling both the galaxy auto-correlation and its cross-correlation with the tracer $U$ as a linear combination of three terms: a linearly-biased component, a white noise-like component, and a scale-dependent bias component proportional to $k^2$. The amplitudes of the linearly-biased contributions for both $P_{gg}(k)$ and $P_{gU}(k)$, extracted using only scales $k<k_{\rm max}=0.3\,{\rm Mpc}^{-1}$, can then be combined to obtain a reliable measurement of $\bU$ that is independent of the small-scale clustering of the galaxies used. We have also shown that the recovered constraints on $\bU$ are robust against uncertainties in the template used to describe the matter power spectrum, and simple parametrisations such as \hfit are sufficient to obtain accurate measurements. Furthermore, we have demonstrated that the dependence of the estimated $\bU$ on the cosmological model chosen to generate the matter power spectrum template can be taken into account through a simple rescaling of the measurements (see Section \ref{sssec:res.val_bm.cosmo}).

  Applying this estimator to cross-correlations between galaxies and maps of the electron thermal pressure and the star-formation rate density, we have shown that unbiased measurements of $\bPe$ and $\brsfr$ can be obtained. Furthermore, we have shown that, while these measurements can be accurately interpreted within the halo model framework, the reliability of the halo model predictions depends critically on the accuracy with which its ingredients (namely halo abundances, linear halo bias, and a physical model describing the halo-observable relation) may be parametrised. Alternatively, accurate predictions may be obtained given a model for the matter-matter and matter-$U$ power spectrum, derived from either simulations or first principles.

  The results presented here should be broadly applicable to the tomographic reconstruction of other astrophysical quantities, such as the cosmic radio \citep{2311.17641}, infrared \citep{2504.05384}, ultraviolet \citep{1810.00885}, gamma-ray \citep{2307.14881}, neutrino \citep{2405.09633}, and gravitational wave \citep{2406.19488} backgrounds. This estimator could also be applied to the study of CMB lensing tomography, as an alternative approach to reconstruct the growth history independently of the small-scale galaxy bias. Finally, the same principles could be applied within the clustering redshifts approach, using the cross-correlation of spectroscopic and photometric galaxies, to reconstruct the bias-weighted redshift distribution of the latter.

\section*{Acknowledgements}
  We would like to thank Raul Angulo, and Jaime Salcido for useful comments. SMM, MZ, and DA acknowledge support from the Beecroft Trust.
  
  This work used the DiRAC@Durham facility managed by the Institute for Computational Cosmology on behalf of the STFC DiRAC HPC Facility (\url{www.dirac.ac.uk}). The equipment was funded by BEIS capital funding via STFC capital grants ST/K00042X/1, ST/P002293/1, ST/R002371/1 and ST/S002502/1, Durham University and STFC operations grant ST/R000832/1. DiRAC is part of the National e-Infrastructure.

  This work was also supported by the Science and Technology Facilities Council (grant number ST/Y002733/1)

\section*{Data Availability}
  Any power spectrum measurement used for this paper and not already available in the \flamingo data release\footnote{\url{https://flamingo.strw.leidenuniv.nl/power_spectra.html}} can be provided upon request.



\bibliographystyle{mnras}
\bibliography{PAPER_v1} 



\appendix

\section{Parameter inference}\label{app:like}
  This appendix describes, in detail, the procedure used to obtain estimates of $\bU$ from the auto- and cross-correlation measurements extracted from the simulations.

  Let ${\bf d}\equiv\left(\hat{P}_{gg}(k),\hat{P}_{gU}(k)\right)$ be a vector containing the measurements of the galaxy auto-spectrum and the cross-spectrum with another LSS tracer $U$. We will assume these measurements to have been made in a discrete set of $k$ bins, $\vec{k}$ covering a range of scales $k_{\rm min}<k_i<k_{\rm max}$. Let ${\sf C}$ be the covariance matrix of these measurements. For simplicity, we will estimate ${\sf C}$ to be diagonal across different $k$ bins, and to take the form:
  \begin{align}\nonumber
    &{\rm Cov}\left(\hat{P}_{wx}(k_i),\hat{P}_{yz}(k_j)\right)=\\
    &\hspace{60pt}\frac{\delta_{ij}}{N_k}\left[P_{wy}(k_i)P_{xz}(k_i)+P_{wz}(k_i)P_{xy}(k_i)\right],
  \end{align}
  where $N_k$ is the number of independent Fourier modes over which the fields involved have been measured that fall within the corresponding $k$ bin. This estimate of the statistical uncertainties does not represent the actual scatter of the power spectrum measurements made from the FLAMINGO simulations. On the one hand, the expression above is only accurate for Gaussian fields. On the other hand, the initial conditions of the fiducial FLAMINGO simulation used here were generated with fixed amplitudes for modes with $k<0.05\,{\rm Mpc}^{-1}$ to reduce the impact of large-scale cosmic variance. Nevertheless, although the estimate above is likely an under-estimation of the true statistical uncertainties, it should provide a sufficiently accurate scheme to weight the different $k$-modes entering the estimator, as well as the level of correlation between the different power spectra.

  To extract constraints on the free parameters of our model $\vec{\theta}$, we may define the Gaussian log-likelihood
  \begin{equation}
    -2\log p({\bf d}|\vec{\theta})=\chi^2(\vec{\theta})\equiv\left({\bf d}-{\bf t}(\vec{\theta})\right)^T{\sf C}^{-1}\left({\bf d}-{\bf t}(\vec{\theta})\right).
  \end{equation}
  Here ${\bf t}(\theta)$ is the theoretical prediction. Inspecting Equations~\ref{eq:P_gg} and \ref{eq:P_Ug}, we may write ${\bf t}$ as
  \begin{equation}
    {\bf t}(\vec{\theta})={\sf T}\,\vec{\theta},
  \end{equation}
  where the parameter vector is
  \begin{equation}
    \vec{\theta}=(N_{gg},N_{gU},A_{gg}^0,A_{gU}^0,A_{gg}^1,A_{gU}^1,...),
  \end{equation}
  and the template matrix ${\sf T}$ takes the form:
  \begin{equation}
    {\sf T}=\left(
    \begin{array}{ccccccc}
      \vec{1} & \vec{0} & \tilde{P}(\vec{k}) & \vec{0}                 & \vec{k}^2\tilde{P}(\vec{k}) & \vec{0} & \cdots\\
      \vec{0} & \vec{1} & \vec{0}                 & \tilde{P}(\vec{k}) & \vec{0}                          & \vec{k}^2\tilde{P}(\vec{k}) & \cdots
    \end{array}\right),
  \end{equation}
  where $\vec{1}$ and $\vec{0}$ are arrays of ones and zeros with the same size as $\vec{k}$, and $\tilde{P}(\vec{k})$ and $\vec{k}^2\tilde{P}(\vec{k})$ are arrays containing the values of the power spectrum template $\tilde{P}_{mm}(k)$, and of $k^2\tilde{P}_{mm}(k)$ in each $k$ bin. Thus, the first two columns of ${\sf T}$ correspond to the white noise components of $\hat{P}_{gg}$ and $\hat{P}_{gU}$. The next two columns correspond to the linear-bias contributions, with $A_{gg}^0=b_g^2$ and $A_{gU}^0=b_g\bU$. All subsequent columns correspond to the polynomial expansion terms in Equations~\ref{eq:P_gg} and \ref{eq:P_Ug}, aimed at absorbing the impact of non-linear biasing.

  Since the log-likelihood is a quadratic function of the parameters, we can obtain analytical estimates of the best-fit parameter values $\vec{\theta}_{\rm BF}$ and their statistical uncertainties (i.e. their covariance ${\sf C}_\theta\equiv\langle (\vec{\theta}-\vec{\theta}_{\rm BF})(\vec{\theta}-\vec{\theta}_{\rm BF})^T\rangle$). These are given by:
  \begin{equation}
    {\sf C}_\theta^{-1}={\sf T}^T{\sf C}^{-1}{\sf T},\hspace{12pt}
    \vec{\theta}_{\rm  BF}={\sf C}_\theta\,{\sf T}^T{\sf C}^{-1}{\bf d}.
  \end{equation}
  Once $\vec{\theta}_{\rm BF}$ is calculated, the values of $b_g$ and $\bU$ can be estimated from the best-fit amplitudes $A_{gg}^0$ and $A_{gU}^0$ as
  \begin{equation}
    b_g=\sqrt{A_{gg}^0},\hspace{12pt}\bU=\frac{A^0_{gU}}{\sqrt{A_{gg}^0}}.
  \end{equation}

  The uncertainties on the measured $b_g$ and $\bU$ can also be calculated analytically in terms of the covariance of $\vec{\theta}$, using linearised error propagation, assuming the relative error of the estimated $b_g$ is small:
  \begin{align}
    &{\rm Var}(b_g)=\frac{{\rm Var}(A_{gg}^0)}{4A_{gg}^0},\\\nonumber
    &{\rm Var}(\bU)=\frac{{\rm Var}(A_{gU}^0)}{(A_{gg}^0)^2}+\frac{(A_{gU}^0)^2}{4(A_{gg}^0)^4}{\rm Var}(A_{gg}^0)\\
    &\hspace{50pt}-\frac{A_{gU}^0}{(A_{gg}^0)^3}{\rm Cov}(A^0_{gg},A^0_{gU}),\\
    &{\rm Cov}(b_g,\bU)=\frac{{\rm Cov}(A_{gg}^0,A_{gU}^0)}{2A_{gg}^0}-\frac{A_{gU}^2}{4(A_{gg}^0)^2}{\rm Var}(A_{gg}^0).
  \end{align}
  This approximation can be easily avoided if necessary. To do so, we may simply treat the measured values of $A^0_{gg}$ and $A^0_{gU}$ as data, fitting them to a model with free parameters $b_g$ and $\bU$ using a Gaussian likelihood with a covariance matrix constructed from ${\sf C}_\theta$. Since this is a simple 2D model, the resulting likelihood is easy to sample efficiently through a variety of methods. We find that the approximate expressions above are sufficiently accurate for all the cases explored in this paper.

\section{Importance on the accuracy of $b_h$}\label{app:bh}
  This appendix contains extra figures showing the differences between halo-bias models as a function of redshift and halo-mass bins explored in this paper, in connection with the the results in Section~\ref{ssec:bh}.

\begin{figure*}
    \begin{subfigure}{\linewidth}
    \centering
    \includegraphics[width=0.33\linewidth]{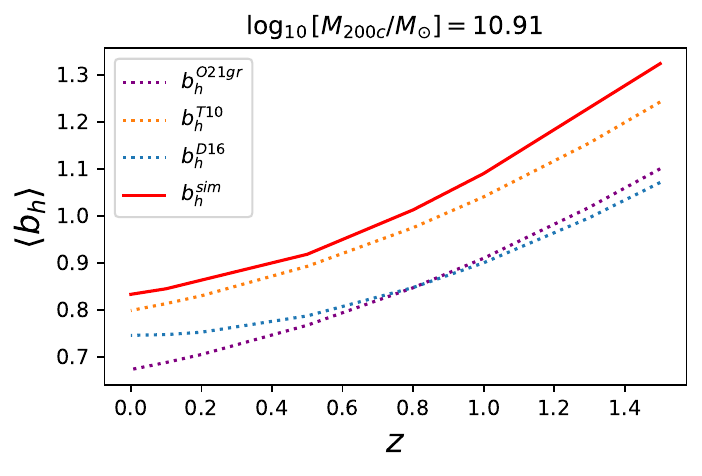}
    \hfill
    \centering
    \includegraphics[width=0.33\linewidth]{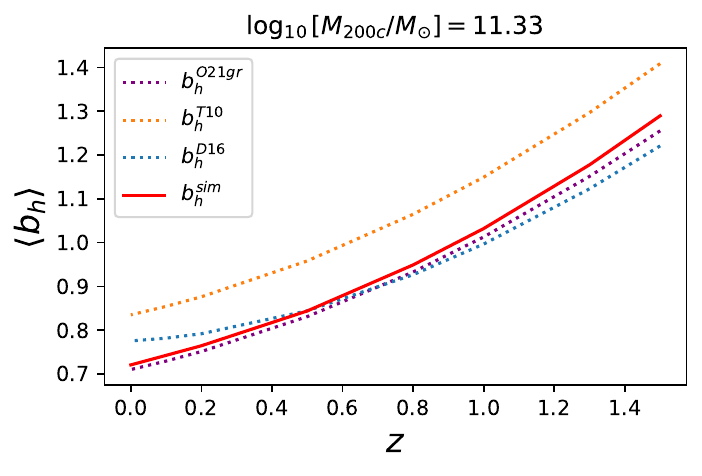}
    \hfill
    \centering
    \includegraphics[width=0.33\linewidth]{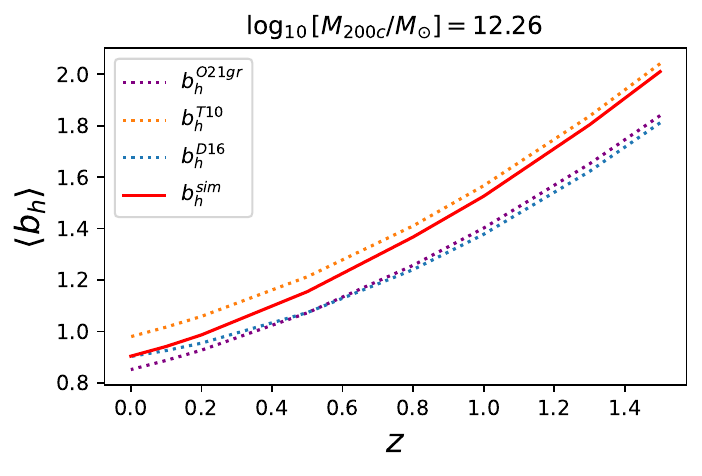}
    \end{subfigure}
    \begin{subfigure}{0.67\linewidth}
    \centering
    \includegraphics[width=0.497\linewidth]{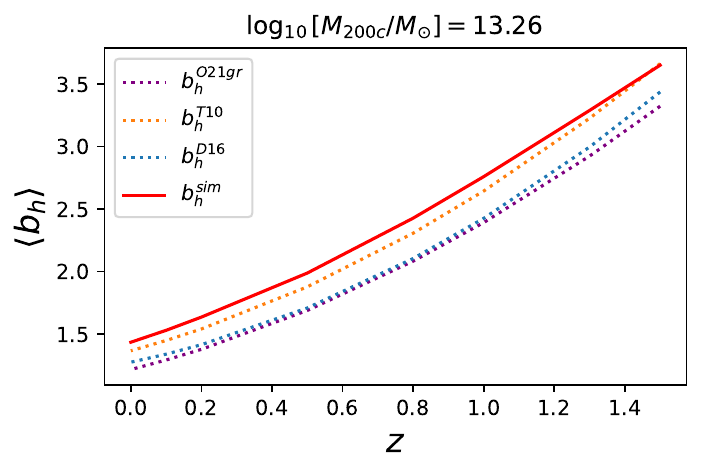}
    \hfill
    \centering
    \includegraphics[width=0.497\linewidth]{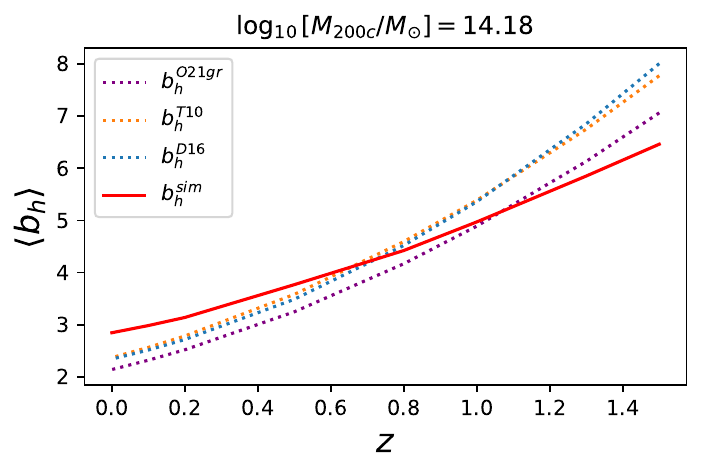}
    \end{subfigure}
    \caption{Halo-bias results from \citet{Tinker10} - blue -, \citet{Despali16} - orange -, and \citet{Ondaro21} - purple -, as well as the result estimated directly from the simulation following an adapted version of Equations~\ref{eq:P_gg} and \ref{eq:P_Ug} - red -. We show the values of $\langle b_h \rangle$ as a function of redshift for the 5 mass-bins studied in the main text (and label with the value of the median mass of each bin)}
    \label{fig:bh}
\end{figure*}

  Figure~\ref{fig:bh} shows the differences in the value of the halo-bias $\langle b_h \rangle$, estimated from 3 different models in the literature \citep{Tinker10, Despali16, Ondaro21} as well as measured directly from the simulation, for the 5 different halo-mass bins explored in this paper. We can see that there is no consistent result among them, and that their variations are in line with those observed in Figure~\ref{fig:bU_Sim_vs_HM}. Given the fact that we cannot really justify the use of one model over the other, we decided to apply our method for estimating $\bU$ to $\langle b_h \rangle$, and use the measured bias from the simulation directly in our analysis. As seen in the main text, this is also the prescription leading to the best results.

\bsp	
\label{lastpage}
\end{document}